\documentclass[aps,showpacs,prl,twocolumn,preprintnumbers]{revtex4}
\begin{document}

\preprint{{DOE/ER/40762-294}\cr{UMPP\#04-003}}

\count255=\time\divide\count255 by 60 \xdef\hourmin{\number\count255}
  \multiply\count255 by-60\advance\count255 by\time
 \xdef\hourmin{\hourmin:\ifnum\count255<10 0\fi\the\count255}

\newcommand{\xbf}[1]{\mbox{\boldmath $ #1 $}}

\newcommand{\sixj}[6]{\mbox{$\left\{ \begin{array}{ccc} {#1} & {#2} &
{#3} \\ {#4} & {#5} & {#6} \end{array} \right\}$}}

\newcommand{\threej}[6]{\mbox{$\left( \begin{array}{ccc} {#1} & {#2} &
{#3} \\ {#4} & {#5} & {#6} \end{array} \right)$}}

\title{Partners of the $\Theta^+$ in Large $N_c$ QCD}

\author{Thomas D. Cohen}
\email{cohen@physics.umd.edu}

\affiliation{Department of Physics, University of Maryland, College
Park, MD 20742-4111}

\author{Richard F. Lebed}
\email{Richard.Lebed@asu.edu}

\affiliation{Department of Physics and Astronomy, Arizona State
University, Tempe, AZ 85287-1504}

\date{September, 2003}

\begin{abstract}
A strangeness +1 exotic baryon $\Theta^+$ has recently been seen in a
number of experiments. We demonstrate that in large $N_c$ QCD the
existence of such a state implies the existence of $S
\! = \! +1$ partner states with various spins and isospins but comparable
masses. We discuss the spectroscopy of such states and possible
channels in which they can be observed, based on the simple assumption
that those states with pentaquark quantum numbers are unlikely to be
large $N_c$ artifacts.
\end{abstract}


\maketitle

The recent experimental observation announced by several
groups~\cite{LEPS,DIANA,CLAS,SAPHIR} of a strange\-ness $S \! = \! +1$
baryon $\Theta^+$ with a mass 1540 MeV and narrow width ($<25$ MeV)
into the channel $K N$ ranks among the most exciting findings in
hadronic physics in recent years.  An $S \!  = \!  +1$ state
necessarily contains an $\bar s$ valence quark, whereas all previously
known baryons have quantum numbers that can be accommodated by three
quarks and no antiquarks.  Given the $K N$ decay channel [as distinct
from $\overline{K} N$, in which conventional $S
\! = \! -1$ resonances such as $\Lambda(1405)$ occur], the most
natural valence content is that of a pentaquark $u u d d \bar s$
state, an entirely new type of hadron.

The width of the $\Theta^+$ has thus far only an experimental
upper bound.  While its small size may at first glance seem
surprising, it is not an uncommon feature among the lowest
strongly-decaying strange baryon resonances ({\it e.g.,} $\Gamma
[\Lambda(1520)] = 16$~MeV), and can be largely attributed to the
smallness of available phase space~\cite{Nus}.

Exotic baryon states were studied previous to their observation, with
some studies appearing as early as the late
1970s~\cite{early,oldskyrme,DPP}---after all, there is no compelling
reason that such states should not occur in QCD.  The recent
announcements by the experimental groups have spawned a flurry of
theoretical work~\cite{quark,soliton,other} using such tools as quark
models with bags, potentials, and pure group theory, and chiral
soliton models.  Nevertheless, all theoretical approaches used up to
this point are more or less model dependent; in this Letter we obtain
model-independent predictions, based upon the existence of the
$\Theta^+$, using only the large $N_c$ limit of QCD.

We should note at the outset that there is no known model-independent
way to predict $\Theta^+$ properties directly from large $N_c$ QCD.
Indeed, claims that chiral soliton model predictions are independent
of model details such as the profile function are shown in
Ref.~\cite{coh} to be the result of a treatment of collective
quantization that is inconsistent with large $N_c$ scaling.

However, large $N_c$ analysis allows one to {\em correlate\/}
predictions of exotic states.  Thus, while large $N_c$ analysis by
itself cannot predict the $\Theta^+$ mass, it {\em can\/} predict the
existence of other $S \! = \! +1$ states with similar masses and
widths ({\it i.e.}, which differ by an amount of order $1/N_c$).  The
quantum numbers of such states are derived here.

In the generalization of QCD from 3 to $N_c$ colors, the ground-state
band of baryons fills a completely symmetric spin-flavor
representation that subsumes the old SU(6) {\bf 56}-plet containing
the $N$, $\Delta$, $\Omega$, and so on~\cite{DJM1}.  Such states have
masses of $O(N_c^1)$, because $N_c$ valence quarks are required to
build a color-singlet state.  Baryons within this multiplet with the
same number of strange quarks are split in mass only at
$O(1/N_c)$~\cite{Jenk}; indeed, the $\Delta$ has enough phase space
for strong decays only because chiral symmetry makes the $\pi$ mass
smaller than the $O(1/N_c)$ $\Delta$-$N$ mass splitting.

Excited baryons exist as well in large $N_c$~\cite{exc,PY}.  While
such resonant states strictly speaking appear as poles in meson-baryon
scattering amplitudes, the consistency between this picture and that
of excited baryons as $N_c$-quark states collected into
SU(6)$\times$O(3) representations in the large $N_c$ limit has been
thoroughly demonstrated~\cite{CL}.  Generically, the well-known Witten
$N_c$ power counting~\cite{Witten} predicts excited baryons to have
widths of $O(N_c^0)$ and masses above those of the ground-state band
by $O(N_c^0)$. The possibility that certain baryons (those in a
mixed-symmetric spin-flavor representation) are actually
characteristically narrower---with widths of $O(1/N_c)$---has been
discussed~\cite{PY}.  However, it has been recently demonstrated that
this result is a consequence of the very specific simple model used;
generically in large $N_c$ QCD the excited baryon widths are, in fact,
$O(N_c^0)$~\cite{CDLN}.  With the inclusion of a typical hadronic
scale $\Lambda_{\rm QCD}$, these excitations typically amount to a few
hundred MeV.

We assume, as is typically done in large $N_c$ studies, that all
quantized observables associated with baryons ({\it i.e.}, spin,
isospin, and strangeness) retain their $N_c \! = \!  3$ values for
arbitrary $N_c$.  Otherwise, one is faced with phenomenological
consequences for $N_c \! > \! 3$ that do not match those of $N_c
\! = \!  3$.   In this spirit, here we assume that the
appropriate large $N_c$ generalization of the $\Theta^+$ is a
state with the quantum numbers of ($N_c \! + \! 1$) light valence
quarks and one valence strange antiquark; the additional light
quarks in the large $N_c$ world form isosinglet, spin-singlet $ud$
pairs.

There is an important distinction between exotic meson and baryon
states at large $N_c$.  For mesons as $N_c \rightarrow \infty$ we know
that there are no narrow $q q \overline{q} \overline{q} $ exotic
states~\cite{Col}, but there must be narrow hybrid exotics with the
quantum numbers of $q \overline{q} g$~\cite{coh-hy}.  For baryons,
large $N_c$ neither implies nor precludes the existence of exotic
states. However, such states fall into nearly degenerate multiplets at
large $N_c$ in a manner analogous to the multiplet structures that
arise for nonexotic baryons ~\cite{Sco,CL}.  Thus, once the existence
of just one such state is established, the existence of a number of
others with different values of spin and isospin is a guaranteed. The
physics underlying this result is the existence of a small number of
``reduced'' scattering amplitudes, each of which contributes to a
number of observable scattering amplitudes~\cite{MP,MM}.  A complex
pole appearing in one of the reduced amplitudes indicates the presence
of a resonant state, with mass and width given by the real and
imaginary parts, respectively.  Furthermore, poles with these same
values then appear in several partial waves, indicating degenerate
states in the large $N_c$ limit.  In fact, since the scattering
amplitudes themselves are $O(N_c^0)$, the masses and widths are
degenerate to this order, and are split only at $O(1/N_c)$, with
typical sizes for $N_c \! = \! 3$ of $<$100~MeV.

We are interested in the appearance of such related resonant states in
$K N$ scattering amplitudes.  These states have $S \! = \! +1$ and are
manifestly exotic.  Since strangeness plays an essential role in this
problem, it may seem natural to work in an SU(3) flavor-symmetric
framework.  But we avoid this for a number of reasons, both practical
and theoretical. On the practical side, our purpose is to predict
exotic states that may be identified experimentally.  $S \! = \! +1$
provides a clean experimental signature of the exotic nature of the
state, and thus we focus on the $S \! = \! +1$ states.  One does not
require a description of full SU(3) multiplets to study these states,
but only the SU(2) multiplets in the $S \! = \! +1$ subspace.  On the
theory side, a number of issues suggest that it is sensible to
restrict ourselves to SU(2) flavor.  For one, SU(3) breaking is not
manifestly small for all observables---while a perturbative treatment
around an SU(3)-symmetric theory works well for many observables, it
also fails for some ({\it e.g.}, the vector meson mass spectrum).  We
do not know {\it a priori} how well it can be expected to work for
exotic baryons, as no one has prior experience with such states; thus,
it seems prudent to refrain from relying upon SU(3) symmetry.
Furthermore, there are subtleties associated with SU(3)
representations at large $N_c$; the SU(3) representations for baryons
are all infinite dimensional as $N_c \! \rightarrow \!
\infty$~\cite{DJM1,DJM2}.  Thus the association of representations at
large $N_c$ with representations at $N_c \! = \! 3$ is not totally
trivial.  In particular, if one inserts $N_c \! = \! 3$ in one part of
the calculation in order to get the physical representations, one
loses the ability to track the $N_c$~\cite{coh}.  While it is possible
to formulate carefully the SU(3) problem at large $N_c$, it is less
ambiguous and more physically transparent to avoid these problems by
imposing only SU(2) isospin symmetry.

Consider meson-baryon scattering $m + B \to m^\prime + B^\prime$ with
fixed strangeness in the initial and final state, such that the meson
$m$ ($m^\prime$) has spin $s$ ($s^\prime$) and isospin $i$
($i^\prime$).  The baryon $B$ ($B^\prime$) belongs to the ground-state
band ($N$, $\Delta$, etc.), which contains for strangeness 0 only
states with spin $\! = \!$ isospin $R$ ($R^\prime$).  The total {\em
spin\/} angular momentum of the meson-baryon system is denoted $S$
($S^\prime$) (and should not be confused with strangeness), while the
relative orbital angular momentum is denoted $L$ ($L^\prime$).  The
total isospin and angular momentum of the state are denoted by $I$ and
$J$, respectively.  Finally, abbreviate the multiplicity $2X+1$ of an
SU(2) representation of quantum number $X$ by [$X$].  The fundamental
expression relating scattering amplitudes in the large-$N_c$ limit
then reads~\cite{MM}
\begin{eqnarray}
S_{L L^\prime S S^\prime I J} & = & \sum_{K, \tilde{K} ,
\tilde{K}^\prime} [K]
([R][R^\prime][S][S^\prime][\tilde{K}][\tilde{K}^\prime])^{1/2}
\nonumber \\
& & \times \left\{ \begin{array}{ccc}
L & i & \tilde{K} \\
S & R & s \\
J & I & K \end{array} \right\}
\! \left\{ \begin{array}{ccc}
L^\prime & i^\prime & \tilde{K}^\prime \\
S^\prime & R^\prime & s^\prime \\
J & I & K \end{array} \right\}
\tau_{K \tilde{K} \tilde{K}^\prime \! L L^\prime} . \nonumber \\
\label{Mmaster}
\end{eqnarray}

A few words about the derivation of Eq.~(\ref{Mmaster}) are in order.
This equation was originally derived in the context of an SU(2)
Skyrme-type model ({\it i.e.}, a model without a strange degree of
freedom), and with the help of standard identities in SU(2) group
theory was then used to deduce the $I_t \! = \! J_t$ rule for
meson-baryon scattering~\cite{MM}. As noted in Ref.~\cite{CL}, such a
derivation, can be turned on its head: The fact that large $N_c$
consistency rules can be shown to imply the $I_t \! = \! J_t$ rule for
all observables at large $N_c$~\cite{KapMan}, together with the same
SU(2) identities, implies that Eq.~(\ref{Mmaster}) holds at leading
order in $1/N_c$.  Such a derivation is fully model independent.
Moreover, this alternative derivation makes clear that
Eq.~(\ref{Mmaster}) applies to the scattering of strange mesons off
nonstrange baryons. The point is simply that the derivation depends
only on the quantum numbers exchanged in the $t$ channel; these
quantum numbers are necessarily non-strange for such a scattering
since the strange quark both enters and leaves the reaction in the
meson, without being transferred to the baryon.  On the other hand,
resonances in the $s$ channel then have $S \! = \! +1$, allowing one
to make contact with the exotic states of interest.

In the present case the mesons are kaons and thus $s=s^\prime=0$,
which collapses the $9j$ symbols to $6j$ symbols, forcing as well $S
\! \to \! R$, $S^\prime \! \to \! R^\prime$, and $\tilde K \! = \!
\tilde{K}^\prime \! = \! K$.  The reduced amplitudes may then be
relabeled $s^{\cal K}_{KLL^\prime} \! = \! (-1)^{L-L^\prime}
\tau_{KKKLL^\prime}$ (${\cal K}$ denotes the kaon).  Furthermore,
$i=i^\prime=1/2$.  In this case the expression simplifies to
\begin{eqnarray}
S_{LL^\prime RR^\prime IJ}
& = & [R]^{1/2} [R^\prime]^{1/2} (-1)^{R-R^\prime} \nonumber \\ & &
\sum_K [K]
\left\{
\begin{array}{ccc} \frac 1 2 & L & K \\ J & I & R \end{array}
\right\}
\left\{
\begin{array}{ccc} \frac 1 2 & L^\prime & K \\ J & I & R^\prime
\end{array}
\right\} s^{\cal K}_{KLL^\prime}. \nonumber \\ & & \label{kmaster}
\end{eqnarray}
All of the predictions of this work follow simply from this relation.

Each distinct pole occurring in Eq.~(\ref{kmaster}) is identified
solely by the $K$ value of the reduced amplitude; the $L$, $L^\prime$
values refer to the manner from which meson-baryon scattering states
couple to the resonances but do not characterize the resonance itself.

For the purposes of this work, we assume that the observed $\Theta^+$
is an $I \! = \! 0$ state.  At present there is no particular
experimental evidence for this value.  However, it does emerge as the
lowest state in many models.  In any case, we will take this as a
starting point for the analysis here both because it is quite
plausible and because the analysis is particularly simple. If it turns
out subsequently that the $\Theta^+$ has a different isospin, the
analysis can be easily modified to account for the correct value.

The isospin of the $\Theta^+$ is not the only unknown quantum number;
its spin and parity also have not been fixed experimentally. Our
prediction of partner states with given quantum numbers depends upon
these values for the $\Theta^+$.  Once they are fixed from experiment,
one can make a concrete predictions.  In the following we make
predictions for partner states based on all of the possible values
$J_0^{P_0}$ consistent with an $N_c \! = \!  3$ pentaquark.  Note that
Eq.~(\ref{kmaster}) with $I \! = \! 0$ is exceptionally simple: The
only surviving amplitudes are
\begin{equation} \label{Ieq0}
S_{L_0 L_0 \frac 1 2 \frac 1 2 0 J_0} = s^{\cal K}_{J_0 L_0 L_0} .
\end{equation}
In particular, the only allowed $K$ value equals $J_0$, only $R \!  =
\! R^\prime \! = \! \frac 1 2$ (the pole appears in $K N$, but not
$K \Delta$, channels due to $I$ conservation), and $L_0^\prime \! = \!
L_0$ (the pole does not appear in mixed partial waves due to angular
momentum and parity conservation).  From here, the procedure is
extremely straightforward: One looks for channels with distinct
($I$,$J$) values in which poles with $K \! = \! J_0$ occur.

The results obtained from Eq.~(\ref{kmaster}) are straightforward to
summarize.  A state of $I \! = \! 0$ and spin $J_0$ and either parity
$P_0$ occurring in a partial wave of relative orbital angular momentum
$L_0$ has $I \! = \! 1$ partners [meaning masses and widths equal to
within $O(1/N_c)$] of all spins $J_1$ consistent with the vector
addition ${\bf J_1} \! = \! {\bf J_0} \! + \! {\bf 1}$, and the same
parity $P_1 \!  = \! P_0 \! \equiv P$.  Furthermore, the $I \! = \!
1$ partners appear only in partial waves with the same orbital quantum
number, $L_1 \! = \! L_0 \equiv L$, which is constrained by angular
momentum conservation to lie within 1/2 unit of $J_0$.  These latter
two results rely on parity conservation.

The previous scheme lists all the possible partner states for a world
with $N_c \! \to \! \infty$.  Clearly some of these could be large
$N_c$ artifacts.  On the other hand, the $J_0$ values that may be
constructed as true pentaquarks at $N_c \! = \! 3$, namely, $1/2$,
$3/2$, and $5/2$, are physically very plausible.  Moreover, for any $I
\! = \! 0$ state containing a (necessarily equal) number of $u$ and
$d$ quarks, it is possible to construct one of the same quark
content---and hence for the same value of $N_c$---but $I \! = \! 1$ by
flipping relative signs of the $u$-$d$ flavor wave function.  One then
finds, for either value of $P$, that $I \! =
\! 1$ partners with each allowed $J_1$ appear in all channels coupled
to either $K N$ ($R \! = \! 1/2$) or $K \Delta$ ($R \! = \! 3/2$),
with the following exceptions:

$\bullet$ If $J_0^P \! = \! 1/2^-$, then only amplitudes with $J_1
\! = R \! = R^\prime$ contain the pole with $K \! = \! J_0$
({\it i.e.}, $K N \to K \Delta$ does not produce this resonance).

$\bullet$ If either (1) $J_0^P \! = \! 3/2^+$ and $J_1^P \! = \!
5/2^+$, (2) $J_0^P \! = \! 3/2^-$ and $J_1^P \! = \! 1/2^-$, (3)
$J_0^P \! = \! 5/2^-$ and $J_1^P \! = \! 7/2^-$, or (4) $J_0^P \! = \!
7/2^+$ and $J_1^P \! = \! 3/2^+$, then the $K \! = \!  J_0$ pole
appears only in amplitudes with $R \! = R^\prime \! = 3/2$.  In such
cases, the given partners would not be visible in $K N \to K N$ or $K
N \to K \Delta$ processes, and therefore alternate experiments to $K
N$ scattering would be required to uncover the existence of such
partners.

If any of the pentaquark states produced in $K N$ scattering---whose
total spin angular momentum is only 1/2---should have a large spin
such as 5/2 or 7/2, then it must be produced in a high partial wave,
say $L \! = \! 2$ or 3.  Assuming that the mass of the state is near
the $K N$ threshold, as is true for the $\Theta^+$ and therefore also
for its partners (degenerate to within about 100~MeV), then the
available phase space is proportional to $| {\bf p} |^{2L+1}$ and
quite small widths [$O$(1~MeV) or less] for such states are not out of
the question.

Equation (\ref{kmaster}) also predicts $I \! = \! 2$ partners to the
$I \! = \! 0$, spin $J_0$ state in large $N_c$. One can show that for
either parity, partners with each spin allowed by the vector addition
rule ${\bf J_2} \! = \! {\bf J_0} \! + \! {\bf 2}$ occur; in the
context of $K R$ scattering, where again $R$ is a ground-state band
nonstrange baryon with $I \! = \! J$, $I \! = \! 2$ states may only be
reached through the $K \Delta$, and some require scattering with the
$N_c \! \ge \! 5$ baryon with $R \! = \! 5/2$. Nevertheless, $I \!  =
\! 2$ states may be reached through other channels such $\gamma d$.
An important caveat to keep in mind when considering higher-isospin
partners, however, is that such states may be artifacts of $N_c \! >
\! 3$, since their total isospin may include $u$, $d$ valence quarks
beyond the four available in the $N_c \! = \! 3$ pentaquark.

Thus we see, regardless of the spin and parity of the $\Theta^+$, that
large $N_c$ QCD predicts it has quantum-number exotic partners
degenerate in mass and width up to $O(1/N_c)$ effects.  While one does
expect the mass prediction to work fairly well---states within a
couple of hundred MeV---a note of caution should be added about the
widths.  The $\Theta^+$ is rather close to threshold, far closer than
is ``natural'' from $1/N_c$ effects alone.  As a result, phase space
greatly restricts its width.  In contrast, its partner states may be
expected to be significantly higher above threshold, which greatly
increases the phase space.  Accordingly, one does not necessarily
expect the widths of these partner states to be similar to that of the
$\Theta^+$, but similar to each other.

We note lastly that Eq.~(\ref{kmaster}) holds for any fixed value
of $S$ (the strange quarks merely ``go along for the ride''), as
long as the mesons in the scattering process have $i \! = \!
1/2$.  That is, Eq.~(\ref{kmaster}) works equally well for
$\overline{K} N$ scattering.  Our results predicting the partners
of the $\Theta^+$ carry over verbatim to predictions of partners
of the $\Lambda$ resonances.  They imply that a $\Lambda$ with
spin-parity $J_0^P$ appearing in a partial wave of a given $L$
has $\Sigma$ partners with spins satisfying the vector addition
rule ${\bf J_1} \! = \! {\bf J_0} \! + \! {\bf 1}$, in the same
partial wave $L$, with the same parity $P$.  This analysis does
{\em not\/}, however, predict the multiplicity of states with
degenerate $I$, $J^P$ corresponding to the same $K$ pole but
distinguished by quantum numbers not specified here. An excellent
example is provided by the $\Lambda(1405)$ and the
$\Lambda(1670)$, both of which have $I \! = \!  0$ and $J^P \! =
\! 1/2^-$, and both are generally assigned to the mixed-symmetry
spin-flavor multiplet of SU(6) (a {\bf 70} for $N_c \!  = \! 3$),
but the former is an SU(3) singlet and the latter is in an SU(3)
octet for $N_c \! = \! 3$. Equation~(\ref{Ieq0}) indicates that
both states correspond to $K \! = \! 1/2$ poles, but because the
masses can be accommodated by (substantial but unnaturally large)
$O(1/N_c)$ corrections~\cite{exc}, while distinct poles with a
given $K$ are generically expected to be separated by $O(N_c^0)$,
one concludes that they both originate from the {\em same\/} $K
\! = \! 1/2$ pole.

One then concludes that in large $N_c$, regardless of any particular
picture such as the quark model, $\Lambda$ resonances of a given
spin-parity should always have $\Sigma$ partners with quantum numbers
as described above.  This is a testable proposition, for which a quick
survey of the {\it Review of Particle Physics}~\cite{PDG} is
appropriate.  One finds that the $J_0^P \! = \! 1/2^-$ $S_{01}$ states
$\Lambda(1405)$ and $\Lambda(1670)$ appear to have as a partner the
$J^P \! = \! 1/2^-$ $S_{11}$ $\Sigma(1620)$, while the $J_0^P \! = \!
1/2^-$ $S_{01}$ state $\Lambda(1800)$ appears to be partnered with the
$J^P \! = \! 1/2^-$ $S_{11}$ $\Sigma(1750)$.  The $3/2^-$ $D_{03}$
resonances $\Lambda(1520)$ and $\Lambda(1690)$ should be partnered
with $3/2^-$ and $5/2^-$ $\Sigma$'s visible in $\overline K N$
scattering, and indeed there appear to exist $D_{13}$ $3/2^-$ states
$\Sigma(1580)$, $\Sigma(1670)$, and $\Sigma(1940)$, and the $D_{15}$
state $\Sigma(1775)$.  One more example is appropriate: The $1/2^+$
$P_{01}$ states $\Lambda(1600)$ and $\Lambda(1810)$ should be
partnered with $1/2^+$ and $3/2^+$ $\Sigma$'s and indeed there exist
$P_{11}$ $\Sigma(1660)$ and $\Sigma(1880)$, but the evidence for the
$P_{13}$ states below 2~GeV is still poor.  While a number of
predicted $\Sigma$ states have not yet been seen definitively, each
one that has been observed can be identified as being partnered with
some observed $\Lambda$ state.

In summary, the large $N_c$ limit of QCD provides a powerful tool to
determine multiplets of baryon states related by symmetry, even in
interesting cases like that of the $\Theta^+$, where the detailed
dynamics underlying their mere existence remains obscure.

{\it Acknowledgments.}  T.D.C.\ was supported by the D.O.E.\ through
grant DE-FGO2-93ER-40762; R.F.L.\ was supported by the N.S.F.\ through
grant PHY-0140362.


\begin{thebibliography}{99}

\bibitem{LEPS}
LEPS Collaboration (T.~Nakano {\it et al.}), Phys.\ Rev.\ Lett.\ {\bf
91}, 012002 (2003).

\bibitem{DIANA}
DIANA Collaboration (V.V.~Barmin {\it et al.}), hep-ex/\ 0304040.

\bibitem{CLAS}
CLAS Collaboration (S.~Stepanyan {\it et al.}), hep-ex/\ 0307018.

\bibitem{SAPHIR}
SAPHIR Collaboration (J.~Barth {\it et al.}), hep-ex/0307083.

\bibitem{Nus}
It has been argued, however, that the widths reported in the
experiments may be broadened by experimental issues associated with
resolution.  Comparison with previous data seems to suggest that the
actual width may be much narrower.  This argument is detailed in
S.~Nussinov, hep-ph/0307357.

\bibitem{early}
H.~H\o gaasen and P.~Sorba, Nucl.\ Phys.\ {\bf B145}, 119 (1978);
M.~de~Crombrugghe, H.~H\o gaasen and P.~Sorba, Nucl.\ Phys.\ {\bf
B156}, 347 (1979); D.~Strottman, Phys.\ Rev.\ D {\bf 20}, 748 (1979);
C.~Roisenel, Phys.\ Rev.\ D {\bf 20}, 1646 (1979).

\bibitem{oldskyrme}
Y.~Oh, B.Y.~Park, and D.P.~Min, Phys.\ Lett.\ {\bf 331B}, 362 (1994);
D.O.~Riska and N.N.~Scoccola, Phys.\ Lett.\ {\bf 299B}, 338 (1993).

\bibitem{DPP}
D.~Diakonov, V.~Petrov, and M.~Polyakov, Z. Phys.\ A {\bf 359}, 305
(1997).

\bibitem{quark}
Fl.~Stancu and D.O.~Riska, hep-ph/0307010; S.~Capstick, P.R.~Page, and
W.~Roberts, Phys.\ Lett.\ {\bf 570B}, 185 (2003); B.G.~Wybourne,
hep-ph/0307170; A.~Hosaka, hep-ph/0307232; R.L.~Jaffe and F.~Wilczek,
hep-ph/0307341; M.~Karliner and H.J.~Lipkin, hep-ph/0307343;
S.-L.~Zhu, hep-ph/0307345; C.E.~Carlson, C.D.~Carone, H.J.~Kwee, and
V.~Nazaryan, hep-ph/0307396; K.~Cheung, hep-ph/0308176; L.Ya.~Glozman,
hep-ph/0308232; R.D.~Matheus, F.S.~Navarra, M.~Nielsen, R.~Rodrigues
da~Silva, and S.H.~Lee, hep-ph/0309001.

\bibitem{soliton}
M.V.~Polyakov and A.~Rathke, hep-ph/0303138; H.~Walliser and
V.B.~Kopeliovich, hep-ph/0304058; D.~Borisyuk, M.~Faber, and
A.~Kobushkin, hep-ph/0307370; M.~Praszalowicz, hep-ph/0308114;
H.-C.~Kim, hep-ph/0308242.

\bibitem{other}
J.~Randrup, nucl-th/0307042; Ya.I.~Azimov, R.A.~Arndt,
I.I.~Strakovsky, and R.L.~Workman, nucl-th/0307088; nucl-th/0308012;
T.~Hyodo, A.~Hosaka, and E.~Oset, nucl-th/0307105; X.~Chen, Y.~Mao,
and B.-Q.~Ma, hep-ph/0307381; L.W.~Chen, V.~Greco, C.M.~Ko, S.H.~Lee,
and W.~Liu, nucl-th/0308006; P.~Bicudo and G.M.~Marques,
hep-ph/0308073; W.~Liu and C.M.~Ko, nucl-th/0308034; S.I.~Nam,
A.~Hosaka, and H.-C.~Kim, hep-ph/0308313; B.~Jennings and K.~Maltman,
hep-ph/0308286; L.Ya.~Glozman, hep-ph/0309092; W.~Liu and C.M.~Ko,
nucl-th/0309023.

\bibitem{coh}
T.D. Cohen, hep-ph/0309111.

\bibitem{DJM1}
R.~Dashen and A.V.~Manohar, Phys.\ Lett.\ {\bf 315B}, 425 (1993);
R.F.~Dashen, E. Jenkins, and A.V.~Manohar, Phys.\ Rev.\ D {\bf 49},
4713 (1994).

\bibitem{Jenk}
E.~Jenkins, Phys.\ Lett.\ {\bf 315B}, 441 (1993).

\bibitem{exc}
C.D.~Carone, H.~Georgi, L.~Kaplan, and D.~Morin, Phys.\ Rev.\ D
{\bf 50}, 5793 (1994);  J.L.~Goity, Phys.\ Lett.\ {\bf 414B}, 140
(1997); C.E.~Carlson, C.D.~Carone, J.L.~Goity, and R.F.~Lebed,
Phys.\ Lett.\ {\bf 438}, 327 (1998); Phys.\ Rev.\ D {\bf 59},
114008 (1999); C.E.~Carlson and C.D.~Carone, Phys.\ Lett.\ {\bf
441B}, 363 (1998); Phys.\ Rev.\ D {\bf 58}, 053005 (1998); Phys.\
Lett.\ {\bf 484B}, 260 (2000); C.L.~Schat, J.L.~Goity,
N.N.~Scoccola, Phys.\ Rev.\ Lett.\ {\bf 88}, 102002 (2002);
Phys.\ Rev.\ D {\bf 66}, 114014 (2002); Phys.\ Lett.\ {\bf 564B},
83 (2003); D.~Pirjol and C.L.~Schat, Phys.\ Rev.\ D {\bf 67},
096009 (2003).

\bibitem{PY}
D.~Pirjol and T.-M.~Yan, Phys.\ Rev.\ D {\bf 56}, 5483 (1997); {\bf
57}, 5434 (1998);

\bibitem{CL}
T.D.~Cohen and R.F.~Lebed, Phys.\ Rev.\ Lett.\ {\bf 91}, 012001
(2003); Phys.\ Rev.\ D {\bf 67}, 096008 (2003); hep-ph/0306102
(accepted to appear in Phys.\ Rev. D).

\bibitem{Witten}
E.~Witten, Nucl.\ Phys.\ {\bf B160}, 57 (1979).

\bibitem{CDLN}
T.D.~Cohen, D.C.~Dakin, R.F.~Lebed and A.~Nellore, in preparation.

\bibitem{Col}
S.~Coleman, {\it Aspects of Symmetry\/} (Cambridge University
Press, Cambridge 1985), Chapter~8.

\bibitem{coh-hy}
T.D.~Cohen,  Phys.\ Lett.\ {\bf 427B}, 348(1998).

\bibitem{Sco}
N.N.~Scoccola, Phys.\ Lett.\ {\bf 236B}, 45 (1990).

\bibitem{MP}
M.P.~Mattis and M.E.~Peskin, Phys.\ Rev.\ D {\bf 32}, 58 (1985);
M.P.~Mattis, Phys.\ Rev.\ Lett.\ {\bf 56},1103 (1986); {\bf 63}, 1455
(1989); Phys.\ Rev.\ D {\bf 39}, 994 (1989).

\bibitem{MM}
M.P.~Mattis and M.~Mukerjee, Phys.\ Rev.\ Lett.\ {\bf 61}, 1344
(1988).

\bibitem{DJM2}
R.F.~Dashen, E.~Jenkins, and A.V.~Manohar, Phys.\ Rev.\ D {\bf 51},
3697 (1995); E.~Jenkins and R.F.~Lebed, Phys.\ Rev.\ D {\bf 52}, 282
(1995).

\bibitem{KapMan}
D.B.~Kaplan and  A.V.~Manohar, Phys.\ Rev.\ C {\bf 56}, 76 (1997).

\bibitem{PDG}
Particle Data Group (K.~Hagiwara {\it et al.}), Phys.\ Rev.\ D {\bf
66}, 010001 (2002).  Note especially Table 13.4 and the Baryon Summary
Table on p. 56.

\end{thebibliography}
\end{document}